\documentclass[11pt]{article}
\pdfoutput=1
\usepackage{amsmath}
\usepackage{amssymb}
\usepackage{graphicx,bbm,mathrsfs}
\usepackage{nicefrac}
\usepackage{slashed}
\usepackage{bbm}
\usepackage{geometry}
\usepackage{empheq}
\usepackage{stackrel}
\usepackage{ulem}
\usepackage{jheppub}
\usepackage{setspace}
\usepackage{relsize}
\usepackage{empheq}
\usepackage{pifont}
\usepackage{mathtools}
\usepackage{enumitem}
\usepackage{dcolumn}   
\usepackage{bm}        
\usepackage{graphicx,mathrsfs}
\usepackage{nicefrac}
\usepackage{multirow}
\usepackage{color}
\usepackage{mathtools}
\usepackage{makecell}
\usepackage{environ} 
\usepackage{lipsum} 
\usepackage{wasysym}
\usepackage{hhline,colortbl}  
\usepackage{tikz}
\usepackage{graphicx}
\usepackage{tensor}
\usepackage{dsfont}
\usepackage{simpler-wick}
\usepackage{accents}

\usepackage{mdframed}

\usepackage{cancel}

\usepackage{booktabs}
\usepackage{mdframed}
\usepackage{xcolor}
\definecolor{EqFrame}{RGB}{235,245 ,250 }

 \NewEnviron{Smaller11}{
           \scalebox{1.1}{$\BODY$} 
 } 
 \NewEnviron{Smaller08}{
           \scalebox{0.8}{$\BODY$} 
 }

\newcommand{\be}{\begin{equation}}
\newcommand{\ee}{\end{equation}}
\newcommand{\bea}{\begin{eqnarray}}
\newcommand{\eea}{\end{eqnarray}}

\newcommand{\SM}{\textrm{SM}}
\newcommand{\DM}{\textrm{DM}}

\renewcommand\labelenumi{(\roman{enumi})}
\renewcommand\theenumi\labelenumi

\renewcommand{\Im}{\operatorname{Im}}

\usepackage{titlesec}

\titleformat*{\section}{\Large\bfseries}
\titleformat*{\subsection}{\large\bfseries}
\titleformat*{\subsubsection}{\large\bfseries}
\titleformat*{\paragraph}{\large\bfseries}
\titleformat*{\subparagraph}{\large\bfseries}

\makeatletter
\newcommand*{\prodsym}{%
  \DOTSB
  \mathop{
    \mathchoice
      {\rlap{\kern.3em\rotatebox[origin=c]{-90}{}}{\prod}}
      {\vcenter{\rlap{\kern.2em\rotatebox[origin=c]{-90}{}}}{\prod}}
      {\sum}{\sum}
  }\slimits@
}
\makeatother

\makeatletter
\DeclareFontFamily{OMX}{MnSymbolE}{}
\DeclareSymbolFont{MnLargeSymbols}{OMX}{MnSymbolE}{m}{n}
\SetSymbolFont{MnLargeSymbols}{bold}{OMX}{MnSymbolE}{b}{n}
\DeclareFontShape{OMX}{MnSymbolE}{m}{n}{
    <-6>  MnSymbolE5
   <6-7>  MnSymbolE6
   <7-8>  MnSymbolE7
   <8-9>  MnSymbolE8
   <9-10> MnSymbolE9
  <10-12> MnSymbolE10
  <12->   MnSymbolE12
}{}
\DeclareFontShape{OMX}{MnSymbolE}{b}{n}{
    <-6>  MnSymbolE-Bold5
   <6-7>  MnSymbolE-Bold6
   <7-8>  MnSymbolE-Bold7
   <8-9>  MnSymbolE-Bold8
   <9-10> MnSymbolE-Bold9
  <10-12> MnSymbolE-Bold10
  <12->   MnSymbolE-Bold12
}{}

\let\llangle\@undefined
\let\rrangle\@undefined
\DeclareMathDelimiter{\llangle}{\mathopen}%
                     {MnLargeSymbols}{'164}{MnLargeSymbols}{'164}
\DeclareMathDelimiter{\rrangle}{\mathclose}%
                     {MnLargeSymbols}{'171}{MnLargeSymbols}{'171}
\makeatother

\begin{document}

\vspace*{4mm}

\thispagestyle{empty}

\begin{center}

\begin{minipage}{20cm}
\begin{center}
\Huge
\sc
\hspace{-5.5cm}    Holographic Dark Matter
\end{center}
\end{minipage}
\\[30mm]

\renewcommand{\thefootnote}{\fnsymbol{footnote}}

{\large\bf
Sylvain~Fichet$^{\, a}$ \footnote{sylvain.fichet@gmail.com}\,, 
Eugenio~Meg\'{\i}as$^{\, b}$ \footnote{emegias@ugr.es}\,,
Mariano~Quir\'os$^{\, c}$ \footnote{quiros@ifae.es}\,
}\\[12mm]
\end{center} 
\noindent

${}^a\!$ 
\textit{CCNH, Universidade Federal do ABC,} 
\textit{Santo Andre, 09210-580 SP, Brazil}

${}^b\!$ 
\textit{Departamento de F\'{\i}sica At\'omica, Molecular y Nuclear and} \\
\indent \; \textit{Instituto Carlos I de F\'{\i}sica Te\'orica y Computacional,} \\
\indent \; \textit{Universidad de Granada, Avenida de Fuente Nueva s/n, 18071 Granada, Spain}

${}^c\!$  
\textit{Institut de F\'{\i}sica d'Altes Energies (IFAE) and} \\
\indent \; \textit{The Barcelona Institute of  Science and Technology (BIST),} \\
\indent \; \textit{Campus UAB, 08193 Bellaterra, Barcelona, Spain}

\addtocounter{footnote}{-3}

\vspace*{8mm}
 
\begin{center}
  {  \bf  Abstract }
  
\end{center}
\begin{minipage}{15cm}
\setstretch{0.95}

Cold dark matter may be a fluid (or plasma) residing in a strongly-interacting hidden sector, rather than a population of weakly-coupled particles. Such a scenario admits a {holographic}  description in terms of a cosmological braneworld embedded in the linear dilaton {five-dimensional (5D)} spacetime.
In this framework, dark matter originates from the linear dilaton bulk black hole, whose phase we show to be thermodynamically favored at all temperatures.
We present a natural freeze-in mechanism for the production of holographic dark matter, in which the bulk black hole is fed by 
energy leaking from the brane after inflation.
Our  model is characterized by two free parameters, one of which, the position of the black hole horizon, is fixed by the observed dark matter abundance. The remaining parameter, the 5D Planck scale $M_5$, is consistent with all current experimental bounds provided that 
$M_5\gtrsim 3\times 10^5$ TeV.

    \vspace{0.5cm}
\end{minipage}

\newpage
\setcounter{tocdepth}{2}

\tableofcontents  

\vspace{1cm}
\vspace{1cm}

\newpage

\section{Introduction}
\label{sec:introduction}

Observations across all astronomical scales point to  the existence of an unknown fluid permeating our Universe, commonly referred to as dark matter.  The  dominant view is  that dark matter consists of a population of yet-unknown particles. 
A key assumption underlying this picture is that the hidden sector responsible for dark matter is {\it weakly coupled}. 
In this paper, building on our previous works \cite{Fichet:2022ixi, Fichet:2022xol}, we explore an alternative possibility: dark matter may instead be a fluid originating in a {\it strongly-coupled} hidden sector.\,\footnote{The term \textit{plasma} is also employed in the context of strongly-interacting theories. }

Our framework realizes a form of non-particle dark matter.
Such a possibility has strong phenomenological motivation. Despite  decades of increasingly sensitive experimental searches, no convincing signal of a dark matter particle has yet been observed. This depressing result motivates searching for a dark matter candidate that behaves as a pressureless fluid on astronomical scales, while remaining diffuse at microscopic scales to evade scattering with ordinary matter.

Remarkably, a candidate of this type arises naturally in string theory, in the so-called Little String Theory (LST) models \cite{Aharony:1999ks,Kutasov:2001uf}. 
From the thermodynamics of LST, it can be inferred that LST behaves at finite temperature  as a fluid with \textit{vanishing pressure} \cite{Fichet:2023xbu,Fichet:2023dju},  which is precisely the macroscopic behavior we seek. Since LSTs are intrinsically strongly coupled, direct computations are challenging. However, a holographic description of LST has been known for a long time \cite{Aharony:1998ub}. This description is based on the so-called {\it linear dilaton} (LD) spacetime, a simple background that appears in a variety of string-theoretic contexts. Roughly speaking, the relation between LD and LST plays a role analogous to that between AdS and CFT.

While having a string construction provides an excellent theoretical motivation, we do not need  the string machinery to obtain a model of non-particle dark matter. Instead, it is sufficient to work with  a consistent five-dimensional gravitational framework. In the minimal version which is our focus, the Standard Model (SM) fields are localized on a brane embedded in the 5D LD spacetime. This defines the cosmological LD$_5$ braneworld model, which is the framework explored in this work.

In this holographic description of dark matter, the pressureless fluid  originates from a black hole living in the bulk spacetime. Our previous work \cite{Fichet:2022ixi, Fichet:2022xol} already explored this fascinating phenomenon, which has a well-known AdS analog (see e.g. 
\cite{Witten:1998qj,Gubser:1999vj,Maldacena:2001kr}). It is in fact instructive to contrast our setup with the  minimal AdS$_5$ braneworld, namely the RS2 model \cite{Randall:1999vf}. In RS2 cosmology, the bulk black hole gives rise to a dark radiation component in the effective Friedmann equation, that may spoil predictions of Big Bang Nucleosynthesis \cite{Binetruy:1999hy, Gubser:1999vj, Hebecker:2001nv,Langlois:2002ke,Langlois:2003zb}). By contrast, in the LD$_5$ braneworld this unwanted dark radiation contribution is replaced by a desirable dark matter-like component. By changing the bulk geometry, we turn a fundamental drawback of braneworld cosmology into a virtue.

In this work we present the essential cosmological evolution of holographic dark matter. We show that a freeze-in mechanism naturally takes place, driven  by interactions of the SM with bulk gravitons.  
We assume that the inflaton  couples only to SM fields. 
As a consequence, the dark sector has negligible energy density after inflation, i.e. the bulk black hole is inexistent. 
Starting at the reheating time $t_R$, energy leaking from the brane  feeds the bulk black hole, which grows until the emission rate becomes negligible. At later times, the  horizon radius freezes.
This holographic freeze-in mechanism can  be schematically viewed as
\begin{center}
  \includegraphics[width=1.0\linewidth,trim={0cm 9cm 0cm 4cm},clip]{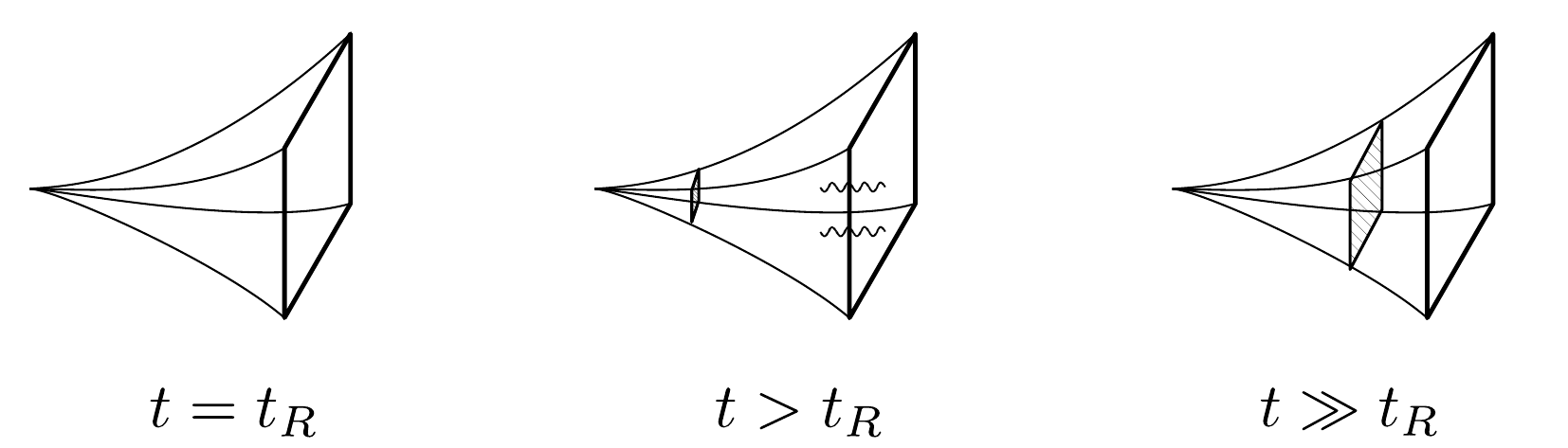}
\end{center}
At late times, the black hole contribution to the brane Friedmann equation simply redshifts as pressureless matter as the universe expands, eventually matching the observed dark matter abundance.

\subsection{Comparison to the Literature}

Let us emphasize the distinctions of the holographic dark matter framework with respect to some other proposals available in the literature. 

First, the holographic  model we propose is unrelated to the recent proposal \cite{Trivedi:2025sbe} in which the behavior of the dark matter energy density is motivated from a bound inspired by the general holographic principle \cite{tHooft:1993dmi,Susskind:1994vu,tHooft:1999rgb,Bousso:1999xy,Bousso:2002ju}. 
In a sense, our model relies on a concrete realization of the holographic principle --- involving the LD$_5$ spacetime.  

Second, although we assume a 4D strongly-interacting dark sector, our proposal is not a version of ``dark QCD''. We do not assume that dark matter consists of a population of dark hadrons, which would bring us back to a  weakly coupled effective field theory description below the (dark) confinement scale. See e.g.~\cite{Fichet:2016clq,Dienes:2016vei,Mitridate:2017oky,Francis:2018xjd,Batz:2023zef} for models in which dark matter is identified with dark hadrons. 
Along the same lines, both flat and warped five-dimensional  scenarios feature weakly-coupled 4D Kaluza-Klein (KK) modes that have been considered as dark matter candidates, see e.g.~\cite{Kong:2005hn,Kakizaki:2006dz,Shah:2006gs,Hooper:2007qk,Belanger:2010yx,Brax:2019koq,Bernal:2020fvw,Kang:2020huh,Kang:2020yul,Cai:2021nmk}.
While our holographic description of dark matter relies on a warped extra dimension, the specific geometry  required (namely LD$_5$) does not produce isolated KK modes. 

Finally, the LD$_5$ geometry that is the essence of our model is also used in the  weakly-interacting continuum proposal of~\cite{Csaki:2021gfm}.
However, here we do not identify dark matter as a population from an infinite, continuous set of  weakly-coupled degrees of freedom, as proposed in \cite{Csaki:2021gfm}.\,\footnote{The 4D degrees of freedom describing such a continuum become strongly-interacting  in the bulk, since  the strength of gravity grows with $r\to 0$ --- notice that the EFT of gravity breaks down at sufficiently  small  $r$, see e.g. \cite{Goldberger:2002cz,Costantino:2020msc}. Also, there is no well-defined  narrow-width approximation for the 4D modes. Instead, the infinite set of modes  has a non-diagonal self-energy matrix, that mixes up all modes \cite{Costantino:2020msc}. 
More generally, a nearly-free continuum is incompatible with standard gravity, see \cite{Fichet:2022xol}.   
See also section \ref{se:comparison} for a related consistency check in our model.}   
Let us mention that a consistent way to realize such a picture is to consider a weakly-coupled discretuum, as done in~\cite{Ferrante:2023fpx}, whose properties are also reminiscent of the Dynamical Dark Matter framework of~\cite{Dienes:2012yz}. The phenomenology and nature of the dark matter in these discretized frameworks are very different from the holographic dark matter model presented here.
In our model, dark matter instead arises from a strongly-coupled sector with no 4D weakly-coupled description.

\subsection{Outline}

This manuscript is organized as follows.  Section~\ref{sec:holographic_fluid} introduces the 5D dilaton-gravity  background with a brane, and presents the properties of the associated holographic fluid. 
 Section~\ref{sec:nu_1}  focuses on a particular case of this background: the linear dilaton. We show that the holographic fluid 
 constitutes a candidate for dark matter, and 
study its thermodynamics. 
 Section~\ref{sec:DM_freezein} presents a natural freeze-in mechanism for holographic dark matter,  that is computed both from the  boundary and bulk perspectives. 
 Section~\ref{sec:phenomenology} presents phenomenological bounds on our model. 
 Section~\ref{sec:conclusions} concludes, and App.\,\ref{app:gravitons} presents details on the linear dilaton graviton propagator and graviton emission.

\section{Holographic Fluids}
\label{sec:holographic_fluid}

In this section we review  a class of  consistent solutions of the brane-dilaton-gravity system, that we refer to as the ${\cal M}_\nu$ spacetimes. We then review the  low-energy 4D effective Einstein equation supported on the brane. In the presence of a bulk black hole, a perfect fluid naturally emerges in the effective Einstein equation, whose  thermodynamics we study.

\subsection{The ${\cal M}_\nu$ Spacetimes }
\label{se:Mnu_sapcetimes}

We consider a 5D spacetime in the presence of a real scalar $\phi$ referred to as the \textit{dilaton} field. The stabilization of the brane-gravity system often involves a flat \textit{brane}. From the low-energy perspective, a brane is simply an infinitely thin hypersurface living in the higher dimensional spacetime and on which operators and degrees of freedom can be localized (see e.g.~\cite{Csaki:2004ay, Sundrum:1998sj,Sundrum:1998ns}).  

We assume the bulk potential
\be
V(\phi)=-\frac{3}{2}(4-\nu^2) k^2 M_5^3 \, e^{2\nu \bar\phi},\quad \bar\phi\equiv\phi/\sqrt{3M_5^3} \,,
\ee
where we introduced the 5D Planck scale  $M_5$. A fully consistent solution of the 5D Einstein field equations in the presence of this potential gives the metric \cite{Fichet:2023dju}
\be
ds^2= g_{MN} dx^M dx^N=e^{-2A(r)}(-f(r)d\tau^2+d\textbf{x}^2)+\frac{e^{-2B(r)}}{f(r)}dr^2   \,,
\ee
with
\begin{align}
A(r)&=-\log(r/L)  \,,  \nonumber\\
B(r)&=-\nu^2\log(r/r_b)+\log(\eta r) \,, \quad \eta = k \, e^{\nu\bar v_b} \,, \\
f(r)&=1-\left(\frac{r_h}{r} \right)^{4-\nu^2} \,,\quad \bar \phi(r)=\bar v_b-\nu \log(r/r_b) \,, \nonumber
\end{align}
with $r_b$ the brane location, and $v_b$ the value of the field $\phi$ at the brane as fixed by a brane-localized  potential. The $\eta$ parameter is a physical scale, {and $L$ is an integration constant with dimension of length that can be fixed to any value.} The $f(r)$ function describes a planar black hole horizon located  at $r=r_h\leq r_b$ in the bulk. See \cite{Fichet:2023xbu} for a careful treatment of the gauge redundancies and integration constants. 

Finally, consistency arguments imply that the allowed range for $\nu$ is $0 \leq \nu <2 $ \cite{Fichet:2023dju,Barbosa:2024pyn}. The $\nu=0$  case corresponds to the AdS$_5$ spacetime --- with a bulk black hole and a brane.

\subsection{Physics on the Brane: Einstein Equation and Holographic Fluids}

The physics felt by a brane-localized observer is governed by the effective Einstein equation 
\begin{equation}
{}^{(4)}G_{\mu\nu} = \frac{1}{M_{4}^{2}} \left( T_{\mu\nu}^b + T_{\mu\nu}^{\rm eff}\right) + O\left( \frac{T_b^2}{M_5^{6}} \right) \,, \label{eq:D_1_Einstein}
\end{equation}
where $T_{\mu\nu}^b$ is the stress tensor of possible brane-localized  matter.
The 4D Planck mass is identified as $ M_4^2 = M_5^3/\eta$. 
The indices in \eqref{eq:D_1_Einstein} are contracted with the  induced metric on the brane, denoted by $\bar g_{\mu\nu}$. Equation \eqref{eq:D_1_Einstein} can be computed in a number of equivalent ways, see e.g.~\cite{Shiromizu:1999wj, Barbosa:2024pyn}.

We are assuming that all components of $T_b$ are small compared to $M_5$. In this low-energy limit, \eqref{eq:D_1_Einstein} takes the form of the standard Einstein equations with an {extra} effective stress tensor $T^{\rm eff}_{\mu\nu}$.  The structure of the $T^{\rm eff}_{\mu\nu}$ tensor is identical to that of a $4$-dimensional {perfect fluid} at rest, 
\begin{equation}
T^{\rm eff,\mu}_\nu = g^{\mu\lambda} T^{\rm eff}_{\lambda \nu} =  \textrm{diag}(-\rho_{\rm eff}, P_{\rm eff}, \cdots, P_{\rm eff}) \,,
\label{eq:Teff_gen}
\end{equation}
that we refer to as the \textit{holographic fluid}. 

What are the properties of this holographic fluid?  Let us choose the induced metric as \be
d\bar s^2=-dt^2+\left(\frac{r_b}{L}\right)^2d\textbf{x}^2  \,,
\label{eq:inducedmetric}
\ee
where the 4D time $t$ and the 5D time $\tau$ are related by $dt= e^{-A(r_b)}\sqrt{f(r_b)} d\tau$.

Tuning the brane tension to remove the effective  4D cosmological constant  hidden in $T^{\rm eff}_{\mu\nu}$, we can simply write $\rho_{\rm eff}\equiv \rho_{\rm fluid}$, $P_{\rm eff}\equiv P_{\rm fluid}$. 
Assuming that  the horizon is very far from the brane, i.e.~$r_h\ll r_b$, we obtain the energy density and pressure 
\begin{align}
\rho_{\rm fluid}&=3\eta^2M_4^2 \left(\frac{r_h}{r_b} \right)^{4-\nu^2} \times \left[1+{\cal O}\left(\frac{r^{4-\nu^2}_h}{r^{4-\nu^2}_b}\right) \right]  \,, \label{eq:rho_fluid} \\ 
P_{\rm fluid}& = \eta^2M_4^2 \left(\frac{r_h}{r_b} \right)^{4-\nu^2} \times \left[ 1-\nu^2 +{\cal O}\left(\frac{r^{4-\nu^2}_h}{r^{4-\nu^2}_b}\right) \right]\,.   \label{eq:P_fluid}
\end{align}

From these results we can infer the equation of state of the holographic fluid as  $P_{\rm fluid}= \omega_{\rm fluid} \,\rho_{\rm fluid}$. Focussing on the leading terms, we find that  $\omega_{\rm fluid} $ is constant with 
 \begin{equation}
w_{\rm fluid} = \frac{1 - \nu^2}{3} \,. \label{eq:w_fluid}
 \end{equation}
 
Remarkably, since $0\leq \nu<2$, $w_{\rm fluid} $ interpolates between the values $\frac{1}{3}$ and $-1$. This is precisely the range of values that has a familiar physical interpretation. For example, $\nu=0$ corresponds to radiation behavior, while $\nu\to 2^-$ corresponds to vacuum energy behavior. 
 
In short, the holographic fluid, at least at rest and for $r_h\ll r_b$,  behaves just like  a familiar perfect fluid, and its equation of state is controlled by the continuous parameter~$\nu$.

\subsection{Thermodynamics of Holographic Fluids}

We proceed with a brief analysis of the thermodynamics. We assume for the moment that the horizon is far from the brane, $r_h\ll r_b$. 

The bulk black hole has a Hawking temperature. This temperature, upon appropriate redshifting, is perceived by a brane observer to be \cite{Fichet:2023dju, Barbosa:2024pyn}~\footnote{The temperature on the brane is obtained from the horizon temperature by multiplying with $1/\sqrt{|g_{\tau\tau}|}$. }
\be
T_b=\frac{4-\nu^2}{4\pi}\eta \left( \frac{r_h}{r_b}\right)^{1-\nu^2} \times \left[1+{\cal O}\left(\frac{r^{4-\nu^2}_h}{r^{4-\nu^2}_b}\right) \right]  \,.  \label{eq:Tb1}
\ee
This temperature is naturally interpreted as the holographic fluid temperature. The higher order corrections come from the effect of the blackening factor contained in the redshifting factor, $1/\sqrt{f(r_b)}$.

Similarly, the bulk black hole  has an entropy density which, once redshifted to the brane observer, is given by
\be 
s_{\rm fluid} =  4\pi \eta M_4^2 \left( \frac{r_h}{r_b} \right)^3 \times \left[1+{\cal O}\left(\frac{r^{4-\nu^2}_h}{r^{4-\nu^2}_b}\right) \right] \,.
\ee 
The free-energy density of the holographic fluid immediately follows: 
\be
f_{\rm fluid}=\rho_{\rm fluid}-T_b s_{\rm fluid}=-\eta^2 M_4^2 \left( \frac{r_h}{r_b}\right)^{4-\nu^2} \times \left[1-\nu^2+{\cal O}\left(\frac{r^{4-\nu^2}_h}{r^{4-\nu^2}_b}\right) \right] \,.  \label{eq:f_fluid1}
\ee

Using this result, we can infer important information about the thermodynamical stability of the fluid. 
We can see that the free energy is negative (positive) for $\nu<1$ ($\nu>1$). 
This implies that the existence of the fluid, i.e.~of the bulk black hole, is thermodynamically favored for $\nu<1$ while it is unfavored for $\nu>1$. Notice however that this conclusion assumes $r_h\ll r_b$. In the $\nu>1$ case, it turns out that the presence  of the fluid becomes favored at larger values of $r_h$ \cite{Barbosa:2024pyn}.  

Finally, for $\nu=1$, the free energy vanishes at leading order. Therefore
the next-to-leading order term must be determined to assess stability. This analysis is carried out in section~\ref{subsec:thermo}.

\section{Dark Matter as the $\nu = 1$ Holographic Fluid}
\label{sec:nu_1}

In this section we interpret the ${\cal M}_\nu$ spacetime as a cosmological braneworld and identify the $\nu=1$ holographic fluid as  a dark matter candidate.

\subsection{Braneworld Cosmology and Friedmann Equations}
\label{subsec:Friedmann_eq}

We assume that all the fields of the Standard Model are localized on the brane. The   $T^b_{\mu\nu}$ stress tensor is identified as the SM stress tensor, $T^b_{\mu\nu} \equiv T^{\SM}_{\mu\nu} $. The total stress tensor in the effective Einstein equation is thus $T^{\textrm{tot}}_{\mu\nu} = T^\SM_{\mu\nu}+ T^{\textrm{fluid}}_{\mu\nu}  $. 

We assume   that the brane is moving, so that $r_b$ depends on time. Within the low-energy regime assumed in \eqref{eq:D_1_Einstein}, the brane motion is nonrelativistic and simplifications occur, see \cite{Fichet:2022ixi, Fichet:2022xol} for details. 
 The effective Einstein equation~(\ref{eq:D_1_Einstein}) then produces  the 
effective Friedmann equations  
\begin{equation}
3 M_4^2 H^2 = \rho_{\textrm{tot}} \,, \qquad  6 M_4^2 \frac{\ddot r_b(t)}{r_b(t)} = - \left( \rho_{\textrm{tot}} + 3 P_{\textrm{tot}}  \right) \,,
\end{equation}
where $r_b(t)/L\equiv a(t)$ plays the role of  the scale factor from the induced  metric~\eqref{eq:inducedmetric}, with $H = \dot r_b/r_b $  the corresponding Hubble parameter. In these equations, we have $\rho_{\textrm{tot}} \simeq \rho_{\textrm{fluid}} + \rho_\SM$,  $P_{\textrm{tot}} \simeq P_{\textrm{fluid}} + P_\SM$. 
The SM fields form a thermal bath with temperature $T_{\rm SM}$, hence  $\rho_\SM=\rho_\SM(T_\SM)$, $P_\SM=P_\SM(T_\SM)$.

The brane position today (i.e. at $t=t_0$) is denoted by $r_b(t_0)=r_{b,0}$.\,\footnote{One could assume $a(t_0)=1$  without loss of generality, in which case $r_b(t_0)=L$. 
We do not make this assumption in the present work. }
Assuming adiabatic expansion of the universe, we have  $\frac{r_b(t)}{r_{b,0}} = \frac{ T_{\SM, 0} }{T_{\SM}(t)}$,  where $T_{\SM,0} = 0.23 $~meV is
the temperature of the universe today.

\subsection{Dark Matter as a Holographic Fluid}
\label{subsec:holo_fluid_DM}

In  the ${\cal M_\nu}$ background with $\nu=1$, the pressure of the holographic fluid vanishes up to ${\cal O}(r_h^3/r_b^3) $ corrections (see \eqref{eq:P_fluid}).
In the context of a cosmological braneworld model, we propose the identification\,\cite{Fichet:2023xbu,Fichet:2022xol} \global\mdfdefinestyle{EqFrame}{ linecolor=white,linewidth=3pt,
backgroundcolor=EqFrame,
leftmargin=3cm,rightmargin=3cm }
\begin{mdframed}[style=EqFrame]
 \begin{center}
 \begin{minipage}{8.5cm}
 \hspace{0.10cm} Holographic fluid $(\nu = 1)$ $\quad\Longleftrightarrow\quad$ Dark Matter
 \end{minipage}
 \end{center}
\end{mdframed}
In the following, we denote  with the subscript ${\rm DM}$ all  quantities related to the holographic fluid.  

At leading order in the $\frac{r_h}{r_b}$ expansion, the thermodynamical properties of our dark matter candidate are 
\be
\rho_{\DM}(t)=3\eta^2 M_4^2 \left(\frac{r_h}{r_b(t)}\right)^3\,,\quad P_{\DM} = 0 \,,\quad f_{\DM}=0  \,,
\label{eq:valores}
\ee
based on (\ref{eq:rho_fluid}), (\ref{eq:P_fluid}), (\ref{eq:Tb1}), (\ref{eq:f_fluid1}). In terms of temperature, we have the scaling $\rho_{\DM}=\rho_{\DM,0} \times \left(\frac{T_\SM}{T_{\SM,0}}\right)^3$. 
Moreover, the DM temperature, i.e.~the Hawking temperature as seen from a brane observer, is approximately constant, and is given by
\be
T_{\rm DM} \simeq \frac{3\eta}{4\pi}  \,.  \label{eq:Tb_DM1}
\ee
The corrections to this picture will be studied in section~\ref{subsec:thermo}.

The constant  temperature of DM is  a hallmark of Hagedorn behavior. 
We remind that the above thermodynamical properties match those of the thermal state of Little String Theory \cite{Aharony:1999ks,Kutasov:2001uf}.  In LST, the Hagedorn temperature is identified as  $T_H=\frac{M_s}{2\pi\sqrt{N}}$~\cite{Fichet:2023xbu}, where $M_s$ is the string scale.

Let us analyze the condition that guarantees that the holographic fluid accounts for the dark matter abundance today. Using today's critical density $\rho_{c,0}= 3 H_0^2 M_4^2$, 
the DM abundance is given by  the remarkably simple formula
\be
 \Omega_{\DM,0}=\left(\frac{\eta}{H_0} \right)^2\left(\frac{r_h}{r_{b,0}}\right)^3   \,. 
\label{eq:rhoDMT}
\ee
Fixing the dark matter abundance to its observed value $\Omega_{\rm DM, 0}\simeq 0.27$ relates therefore the ratio $r_{h}/r_{b,0}$ to the $\eta$ parameter. Using today's value of the Hubble parameter $H_0\simeq 1.47\times 10^{-42}$ GeV, we obtain a bound on $\frac{r_h}{r_{b,0}}$.  
The bound can be extended to any time by considering the adiabatic expansion of the universe $(r_b \propto 1/T_{\SM})$, from which we obtain
\begin{equation}
\frac{r_h}{r_b} \simeq 3.5 \times 10^{-16} \frac{T_{\SM}}{\textrm{GeV}} \left(\frac{\textrm{GeV}}{\eta}\right)^{2/3}   \,.
\label{eq:rh_bound}
\end{equation}
As a result, the horizon is typically  very far from the brane, and the ${\cal O}(r_h^3/r_b^3)$ corrections are tiny, {even for the
maximum reheating temperature $T_{\SM,R}$ that we will find in section~\ref{sec:DM_freezein}.} 

Since the ${\cal M}_\nu$ background has two physical parameters, our dark matter model has two  parameters.    Here we choose them to be  $\{ \eta, r_h\}$. Fixing the DM abundance relates these two parameters as given in \eqref{eq:rhoDMT}. 
A  mechanism for the formation of the black hole horizon, that determines $r_h$ dynamically, will be presented in section \ref{sec:DM_freezein}.

\subsection{Holographic Dark Matter Thermodynamics}
\label{subsec:thermo}

In this subsection we perform a more precise analysis of the thermodynamics  of the holographic  fluid identified as dark matter.

The first law of thermodynamics dictates that the variation of the total energy of a closed system satisfies  $dU = T dS - P dV $ where $U$ and $S$ are extensive quantities.
In particular,  for the holographic fluid at  temperature $T_{\DM}$ we have
\be
dU(r_h,r_b) = T_{\DM}\, dS(r_h,r_b)-P(r_h,r_b)\, dV(r_b) \,,
\ee
which implies
\begin{equation}
\left[ \frac{\partial U}{\partial r_h}  - T_{\DM} \frac{\partial S}{\partial r_h} \right] dr_h + \left[ \frac{\partial U}{\partial r_b} + P \frac{\partial V_b}{\partial r_b} - T_{\DM} \frac{\partial S}{\partial r_b} \right] dr_b = 0  \,.  \label{eq:thermo_relation}
\end{equation}

The exact temperature is 
\be  T_{\DM}(x_h) = \frac{T_H}{f_b^{1/2}} \,, \quad\quad  T_H \equiv \frac{3\eta}{4\pi}   \,, \label{eq:TDM}
\ee
 where $T_H$ is the Hagedorn temperature, and the blackening factor is $f_b(x_h) = 1 - x_h^3$, where 
we introduced  the $x_h=r_h/r_b$ variable. 
  The $T_{\rm DM}$ temperature differs from $T_H$ by 
a ${\cal O}(x_h^3)$ correction, which was neglected in the previous sections.  
  We have $T_{\DM} \geq T_H$. 
  
  
We also  know the exact  free energy density $f_{\rm DM}$, which is the on-shell action (see e.g. \cite{Barbosa:2024pyn}). Using  the free energy and  temperature,  and combining with the thermodynamical relations \eqref{eq:thermo_relation}, we are able to derive the entropy and energy density. The results are 
  \begin{eqnarray}
    \frac{\rho_{\DM}}{\eta^2 M_4^2} &=& 6 \left( 1 - f_b^{1/2} \right) \,, \\
        \frac{s_{\DM}}{\eta M_4^2} &=&  4\pi\left( 1-f_b \right) \,, \label{eq:s_fluid_1}  \\
    \frac{f_{\DM}}{\eta^2 M_4^2} &=& \frac{\rho_{\DM} - T_{\DM} s_{\DM}}{\eta^2 M_4^2} =  - \frac{3}{f_b^{1/2}} \left( 1 - f_b^{1/2}  \right)^2 \,.
    \label{eq:f_fluid}    
  \end{eqnarray}
Expanding these formulas for $x_h\ll1$ reproduces the values in (\ref{eq:valores}). In particular, the free energy density vanishes  up to $\mathcal O(x_h^3)$ corrections, such that $f_{\DM}=\mathcal O(x_h^3)$. 
From this  we can see that $f_{\DM}< 0$ for any value of $r_h>0$. This implies that for $T_{\rm DM}>T_H$ the black hole phase is always  favored with respect to the no-black hole phase.\,\footnote{The situation is different for $\nu>1$, in which case the black hole phase is energetically favored only for $r_h$ above a finite value, see \cite{Barbosa:2024pyn}.  } This result is consistent with the analysis of LST done in \cite{Kutasov:2000jp}.

We can write (\ref{eq:s_fluid_1})-(\ref{eq:f_fluid}) in terms of $T_\DM$ using that
\be
  f_b(T_{\DM}) = \left(\frac{T_H}{T_{\DM}} \right)^2 \,,
\ee
which gives
  \begin{align}
  \frac{s_{\DM}}{\eta M_4^2}= 4 \pi \left( 1 - \frac{T_H^2}{T_{\DM}^2} \right)  \,, \quad 
  \frac{\rho_{\DM}}{\eta^2 M_4^2}= 6 \left( 1 - \frac{T_H}{T_{\DM}}  \right) \,, \quad 
  \frac{f_{\DM}}{\eta^2 M_4^2} = -3 \frac{ \left( T_{\DM} - T_{H}\right)^2  }{T_{\DM} T_H}  \,. \label{eq:fDM}
  \end{align}
These expressions are valid for $T_{\DM} \ge T_H$, with $T_{\DM}$ being typically very close to $T_H$. 
 We can notice the important feature that  $f_{\DM}(T_{\DM})$ and $f_{\DM}^\prime(T_{\DM})$ vanish at  $T_{\DM} = T_H$ while  $f_{\DM}^{\prime\prime}(T_{\DM})$ does not. 

Using these results, it is possible to study the behavior of our thermodynamical system in terms of the brane temperature, $T_{\DM}=T_{\DM}(T_{\rm SM})$. Using Eq.~(\ref{eq:TDM}) and the adiabatic expansion of the universe, we have
\begin{equation}
T_{\DM}(T_{\rm SM}) = \frac{T_H}{\sqrt{1 - \left( \frac{r_h}{r_{b,0}}\right)^3 \left(  \frac{T_{\SM}}{T_{\SM,0}} \right)^3}} \,.
\end{equation}
Notice that the idea of  using thermodynamical variables defined on the boundary of the system has been long known, see
\cite{York:1986it, Hawking:1982dh}. The general principle of our analysis is analogous, except that here the black hole is not in thermal equilibrium with the brane.

The free energy of the black hole phase is given by \eqref{eq:fDM} taken as a function of $T_{\rm DM}$, $f_{\DM}=f_{\DM}(T_{\rm DM })$. The free energy of the phase with no black hole is $f_{\rm  no\,{BH}}(T_{\rm SM })=0$ for all $T_{\rm SM }\geq 0$. Using \eqref{eq:fDM}, we find that $f_{\rm DM}(T_H)=f'_{\rm DM}(T_H)=0$, which implies that there is a continuous  phase transition at $T_{\rm DM}=T_H$. The phase transition being continuous, no supercooling is possible, and   we conclude that the system is in the black hole phase for any $T_{\DM} > T_H$. This applies in particular to the cosmological braneworld model.

Finally we can also compute the DM speed of sound, defined by  $c_\DM^2 \equiv - \frac{\partial f_{\DM}}{\partial \rho_{\DM}} = - \frac{f_{\DM}^\prime(x_h)}{\rho_{\DM}^\prime(x_h)} $, as
   \begin{equation}
  c_{\DM}^2 = \frac{1}{2} \frac{x_h^3}{1 - x_h^3} = \frac{1}{2 T_H^2} \left( T_{\DM}^2 - T_H^2\right)  \,. \label{eq:cDM}
   \end{equation}
 In the present work, $x_h$ is extremely small, i.e. $r_h \ll r_b$, c.f.~\eqref{eq:rh_bound}, so that $c_{\DM}^2 \ll 1$.

One may notice that, when the black hole comes very close to the brane, 
\eqref{eq:cDM} would give $c_\DM > 1$, which is forbidden. This occurs for $T_{\DM} > \sqrt{3} \, T_H$, which corresponds to $r_b \lesssim 1.14 \, r_h$.
 This problem seems to be related to the ambiguity in the definition of the BH vacuum state, as it was argued in Ref.~\cite{Hebecker:2001nv} (see also Ref.~\cite{Page:1982fm}). This extreme regime is not relevant for our study. 

\paragraph{Brief summary.} Our thermodynamical analysis of the $\nu=1$ holographic fluid establishes that 
\begin{itemize}
    \item The system is always in the black hole phase for any $T_{\rm SM}>0$. 
    \item The DM sound speed is nonzero but small, with typically $c^2_\DM\sim \frac{x_h^{3}}{2}$. 
\end{itemize}

\subsection{On the Model Parameters} \label{se:parameters}

Let us discuss the free parameters of the  model. 
The ${\cal M}_{\nu=1}$ background has two free parameters, that can be chosen to be $\{ \eta, r_h\}$. The other parameters $\{M_5$, $T_{\rm DM}\}$ appearing in the model satisfy the relations
\begin{equation}
M_5^3 = \eta M_4^2 \,, \qquad T_\DM = \frac{3 \eta}{4\pi} \frac{1}{\sqrt{1 - (r_h / r_b)^3 }} \,. 
\end{equation}
We remind that $M_4$ is the 4D Planck mass. 
If one neglects $r_h/r_b$, then $T_\DM$ is simply proportional to $\eta$. If the mild dependence in $r_h$ is taken into account, $T_\DM$ becomes an independent parameter that could be traded for $r_h$, although such a choice would be unpractical since it would involve a fine cancelation between $T_\DM$ and $T_H= \frac{3 \eta}{4\pi}$. 

 One combination of our  two free parameters is fixed by the value of DM abundance today obtained from  \eqref{eq:rhoDMT}. Our model has thus a single free parameter.   In the following we choose that $r_h$  is fixed by the DM abundance, while $\eta$ (or equivalently $M_5$) remains as the free parameter. The phenomenological  bounds on  the free parameter are presented in section~\ref{sec:phenomenology}. 
In the next section a dynamical mechanism for DM formation is presented, that does not change the parameter counting.

\section{Freeze-in Production of Holographic Dark Matter}
\label{sec:DM_freezein}

Having identified the $\nu=1$ holographic fluid as  the cosmological dark matter observed  in our universe, we should  ask  about the cosmological history of such a scenario. Is there a mechanism capable of producing the holographic fluid with correct abundance in the early universe?  

In this section we show that a mechanism analogous to freeze-in production~\cite{Hall:2009bx} naturally occurs in our holographic dark matter framework. This is the mechanism  sketched  in section~\ref{sec:introduction}. 
The freeze-in computation  will be done from both the brane and bulk perspectives. Even though the intermediate steps appear different, the final results perfectly match, thereby  demonstrating the self-consistency of the whole formalism.

Our main assumption is that the inflaton is brane-localized. As a result,  the brane position $r_b$ (i.e.~the scale factor) grows exponentially during inflation, such that the bulk can be considered as empty at the end of inflation.  It is thus well-motivated to  consider the  system at the reheating time  with no bulk black hole, i.e.~$r_h=0$. The corresponding reheating temperature is denoted $T_{\SM,R}$.

\subsection{Heating the Bulk: Graviton Radiation from the Brane}
\label{se:graviton_emission}

The Standard Model particles on the brane couple to  bulk gravitons as
\be
\mathcal L \supset -\frac{1}{M_5^{3/2}}T^{\mu\nu}_{\rm SM}(x) h_{\mu\nu}(z_b,x)\,. 
\label{eq:5DLagrangian}
\ee
Therefore,   processes of the form $ \varphi + \bar\varphi \to h_{\mu\nu}$ happen in the brane thermal bath. Since the gravitons live in the bulk, this is an emission mechanism through which energy is transferred from the brane to the bulk.

The rate of this energy transfer process, here denoted $\Delta \dot\rho_\SM$, was estimated in \cite{Gubser:1999vj} and computed in  AdS$_5$ in \cite{Hebecker:2001nv,Langlois:2002ke, Langlois:2003zb}.
We perform a computation of $\Delta \dot\rho_\SM$ in the LD$_5$ background using a unitarity cut on  the LD propagators computed in \cite{Fichet:2023xbu, Fichet:2023dju}. The details of the computation are given in App.~\ref{app:gravitons}. 
For center-of-mass energy much higher than the local 5D curvature $R|_{\rm brane} \approx -12 \eta^2$, the result matches the flat space result (as well as the AdS$_5$ with negligible curvature, as done in \cite{Hebecker:2001nv,Langlois:2002ke, Langlois:2003zb}). 

The brane-to-bulk energy transfer rate is 
\be
\Delta \dot\rho_\SM = -\frac{2\left(\sum_\varphi g_\varphi \kappa_\varphi^2 c_\varphi a_\varphi \right)}{5\pi^4 \eta M_4^2}\Gamma\left(\frac{7}{2}\right)\Gamma\left(\frac{9}{2}\right)\zeta\left(\frac{7}{2}\right) \zeta\left(\frac{9}{2}\right) T_\SM^8  \,, 
\label{eq:deltarho}
\ee
where $a_s=a_V=1,a_f\simeq 0.75$ ($s$, $V$ and $f$ stand for scalars, vectors and fermions, respectively), and $c_\varphi$ is given by Eq.~(\ref{eq:c_varphi}). Considering the SM degrees of freedom with $g_s=4$, $g_V=12$ and $g_f=45$ we get $\sum_\varphi g_\varphi \kappa^2_\varphi c_\varphi a_\varphi = 20.78$, where $\kappa_\varphi$ is the number of spin degrees of freedom of $\varphi$ ($\kappa_s=1$ and $\kappa_V=\kappa_f=2$). Numerically, 
\be
\Delta \dot\rho_\SM = - C_\rho \frac{T_\SM^8}{\eta M_4^2}\,,\quad \quad\quad  C_\rho \simeq 3.919 \,.
\label{eq:c}
\ee

\subsection{Evolution Equations: Boundary Viewpoint}
\label{sec:evolution-boundary}

The 4D effective Einstein equation \eqref{eq:D_1_Einstein} leads to a 4D effective conservation equation. It can be derived either by contracting with $\nabla^\mu$ and using the 4D Bianchi equation, or  by starting with the 5D conservation equation and projecting it on the brane, see \cite{Fichet:2022xol, Fichet:2022ixi} for details and checks. 
The effective conservation equations for the SM and DM energy densities are 
\begin{align}
\dot{\rho}_\SM &=-4 H \rho_\SM + \Delta \dot \rho_\SM \,,   \label{eq:rhob}\\
\dot{ \rho}_\DM &=-3 H \rho_\DM + \Delta \dot \rho_\DM \,, \quad\quad  \Delta \dot \rho_\SM + \Delta \dot \rho_\DM = 0  \,.
\label{eq:rhoDM}
\end{align}

Since by definition $\rho_\DM \propto r^3_h$ (see \eqref{eq:valores}), the evolution equation \eqref{eq:rhoDM} can alternatively be thought as an evolution equation for the  $r_h$ parameter. That is, the equation must describe the evolution of the black hole horizon in the bulk, providing a geometric picture of freeze-in. The evolution equation for $r_h$ is found to be
\be 
9  M_4^2 \eta^2 \frac{\dot r_h r^2_h}{r^3_b} =  \Delta \dot \rho_{\DM}\,. 
\label{eq:rh_evol}
\ee 
In next section we show that this result can be derived directly from the bulk viewpoint.

\subsection{Evolution Equations: Bulk Viewpoint}

\label{sec:evolution-bulk}

The radiation can be modeled using a Vaidya-like metric \cite{Vaidya:1953zza,Lindquist:1965zz}, which describes the spacetime surrounding a radiating relativistic star. 
The underlying approximation  is that the radiation is emitted orthogonally to the brane \cite{Langlois:2003zb}. This approximation holds when the brane is non-relativistic and therefore applies  in the low-energy regime that is our working assumption.

\begin{figure}[t]
\begin{center}
  \includegraphics[width=0.8\textwidth,trim={0cm 3.5cm 0cm 1cm},clip]{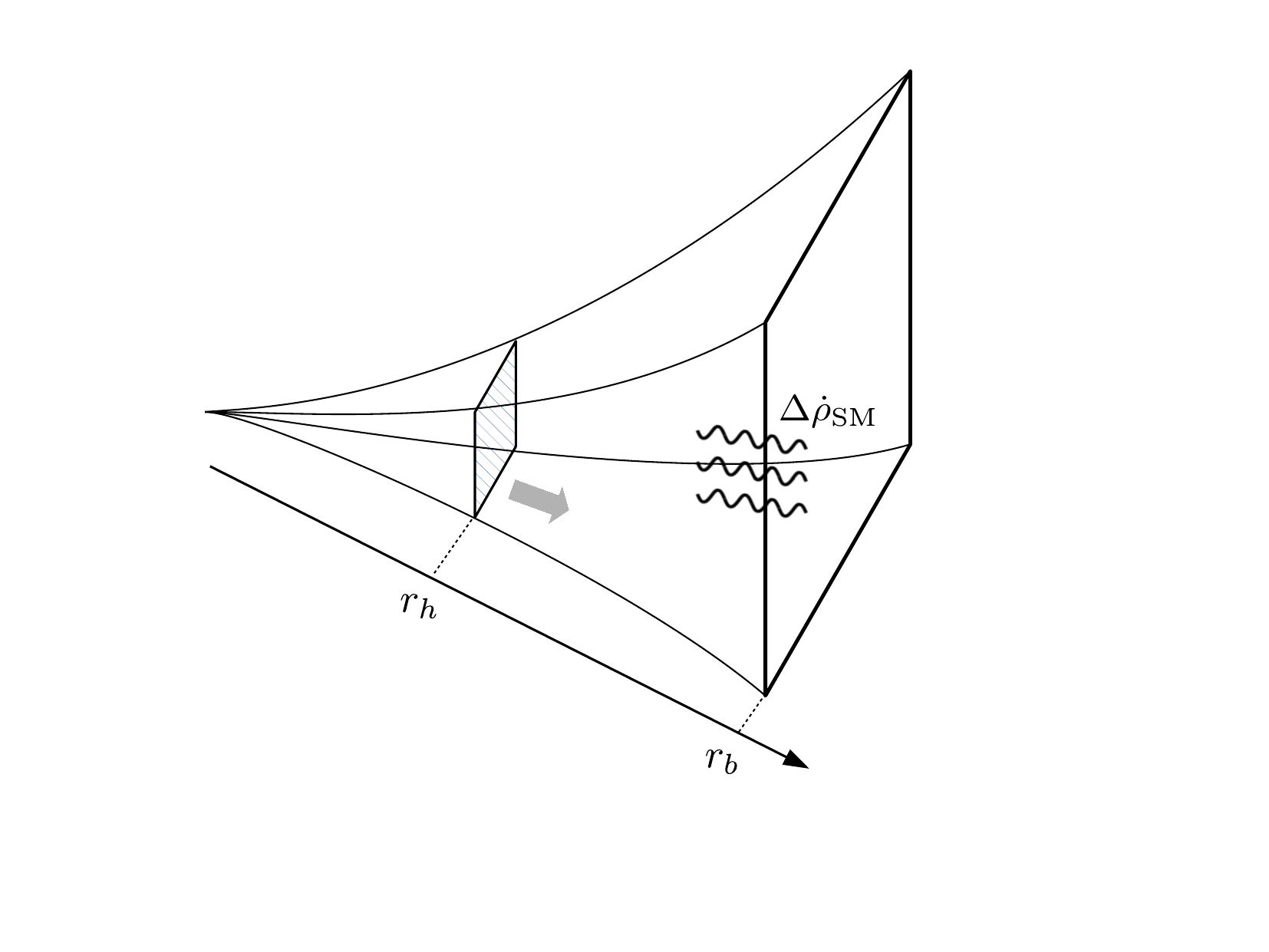}
\end{center}
     \caption{The freeze-in of holographic dark matter.  }
     \label{fig:Freeze_in_full}
 \end{figure}

Along the same lines, we assume that the brane motion is negligible compared to the motion of the bulk horizon, i.e.~$\dot r_h \gg \dot r_b$.  That is, we assume that the freeze-in timescale is much smaller than the universe's expansion. 
The validity of these assumptions will be verified a posteriori. The geometry is summarized in Fig.\,\ref{fig:Freeze_in_full}.

We introduce the light-like coordinate $v$  
\be
v=t + \int \frac{dr}{f(r)}  \frac{L}{r\eta r_b }   \,,
\ee
for which the metric takes the form
\be
ds^2 = - f(v,r) \frac{r^2}{L^2} dv ^2  + 2 \frac{r}{L\eta r_b} dv dr +  \frac{r^2 }{L^2} d{\bm x}^2 \,.
\label{eq:Vaidya_metric}
\ee
In the static  limit for which $r_h$ is constant, i.e. the $f$ factor only depends  on $r$, one can verify that \eqref{eq:Vaidya_metric} is equivalent to the original linear dilaton metric \eqref{se:Mnu_sapcetimes}. 

Using this metric, radiation is described by the bulk stress-energy tensor 
\be
~^{(5)}T^{\rm rad}_{MN} = \Delta \dot \rho_\DM\, k_M k_N  \,,
\ee
where $\Delta \dot \rho_\DM=\frac{d}{dt}\Delta \rho_\SM$ is the energy transfer rate in the brane time, and  $k_M$ is a null vector. $k_M$  is chosen to satisfy  $k_M u^M =1$, with $u^M$  the timelike unit  velocity vector for brane observers,  $u^M = (\dot v,  \dot r,  {\bm 0} ) $.
Using the metric \eqref{eq:Vaidya_metric}, the only non-vanishing component of $k_M$ is $k_t  $ (or $k^r$).  
Moreover,  $k_M u^M =1$ implies  implies $k_t= 1/\dot v$. 
Finally, the normalization  $u^M u_M =- 1$ gives 
\be
\dot v \approx \frac{L}{r \sqrt{f(v,r)}}\,,
\ee
where we used  the nonrelativistic limit for which $\dot r$ can be neglected.\,\footnote{The exact relation is \be
\dot v = \frac{L}{\eta  r_b f r}\left(\dot r  +  \sqrt{\dot r^2 +\eta^2 r_b^2 f }\right) \,.
\ee }

Turning to the Einstein tensor, we find that only the $vv$ component depends on the variation of $r_h$,  $\frac{d}{dv} r_h$. The $vv$ component of the Einstein equation,  $M_5^3 \,^{(5)}G_{vv} = \,^{(5)}T^{\rm rad}_{vv} $, implies
\be
9  L M_4^2  \eta^2 r_b \, r_h^2 \frac{d}{dv} r_h = 2 \Delta \dot \rho_{\DM}  r^5 f \,. \label{eq:EE1}
\ee
We evaluate \eqref{eq:EE1} on the brane,  set $r=r_b$, $f\approx 1$, and use $\frac{d}{dv} = \frac{d}{\dot v \,dt}$. Notice that the $L$ dependence vanishes. The result gives precisely the evolution equation \eqref{eq:rh_evol} --- up to a factor 2 coming from the fact that in the above description, the brane radiates on both sides, as noticed in \cite{Langlois:2002ke}.

\subsection{Initial Condition}

We assume that our holographic model is valid during inflation. 
We further assume that the inflaton field  is brane-localized. Geometrically, this implies that $\dot r_b \propto e^{Ht}$, i.e. the brane runs away exponentially towards positive $r$. As a  result of the  braneworld inflation, we have $r_h\sim 0$,  i.e. the holographic fluid is inexistent at the end of inflation. {In this section we call $r_h(T)$ the value of the horizon as a function of the temperature, as given by sections~\ref{sec:evolution-bulk} and \ref{sec:evolution-boundary}, while its constant value at small temperatures (i.e.~in standard cosmology) will be denoted as $\bar r_h$. } 

In this work we do not specify the inflation mechanism, and simply start at the reheating time, with corresponding temperature $T_{\SM ,R}$ .\,\footnote{Notice however that the scenario is well-motivated by e.g. the trace anomaly driven inflation mechanism of~\cite{Hawking:2000bb}. }
The initial condition for the dark matter energy density is 
\be
\rho_{\DM}(T_{\SM,R} ) =0 \,. 
\label{eq:IC}
\ee

\subsection{Holographic Dark Matter Freeze-In}

In this subsection we solve the  evolution equations \eqref{eq:rhob} and \eqref{eq:rhoDM}. Let us  first notice that, for low temperatures, the energy transfer term  $\Delta \dot \rho_{\SM,\DM}$  is negligible as compared to the other terms in the conservation equations. Because of the initial condition \eqref{eq:IC}, we can 
 also assume that $\rho_\SM\gg \rho_\DM$ throughout the evolution of the Universe, such that 
the Hubble parameter is controlled by the SM thermal bath, 
\be
H=\frac{1}{\sqrt{3}M_4}\sqrt{\rho_\SM +  \rho_\DM}\simeq \frac{1}{\sqrt{3}M_4}\sqrt{\rho_\SM}  \,. \label{eq:H}
\ee
These points imply that, at low temperatures, the  scaling of $\rho_\SM$ and $\rho_\DM$ from standard cosmology is recovered. We denote the corresponding quantities in standard cosmology with bars, 
\be
\bar \rho_\SM (T_{\SM}) \propto T_\SM^4\,, \quad \quad \bar \rho_\DM (T_{\SM}) \propto T_\SM^3   \,.
\ee
Furthermore, we have that $ |\Delta \dot \rho_{\rm SM,R}| \ll H \rho_{\rm SM}(T_{\rm SM,R})$ at reheating. Therefore the  correction  to the SM energy density is small, and we can write 
\be
 \rho_\SM (T_{\SM})   \approx \bar \rho_\SM (T_{\SM}) = g_\star  \frac{\pi^2}{30}T_\SM^4
\ee
to a very good approximation.

Using these simplifications, we now solve the evolution equation~\eqref{eq:rhoDM}. 
The result is 
\begin{align}
\rho_{\rm DM}(T_{\SM}) & =\frac{ \sqrt{10} C_\rho }{\pi\eta M_4\sqrt{g_*}} T_{\SM}^3 \left( T_{\SM,R}^3 - T_{\SM}^3 \right)  \\ 
 & = \bar \rho_{\rm DM}(T_{\SM}) \times \left (1-\frac{T_{\SM}^3}{T_{\SM,R}^3}\right)   \,,
\label{eq:rho_DM}
\end{align}
with 
\be
\bar \rho_{\rm DM}(T_{\SM}) = \frac{\sqrt{10} C_\rho }{\pi\eta M_4\sqrt{g_*}} T_{\SM,R}^3 \,  T_{\SM}^3  \,. \label{eq:rho_DM_result}
\ee
This result constitutes the prediction  for the dark matter energy density in our holographic model. 
From \eqref{eq:rho_DM}, we can see that the freeze-in production occurs within roughly one order of magnitude in temperature below $T_{\rm SM,R}$. 

{The freeze-in production of the dark fluid can equivalently  be interpreted as the freeze-in production of the black hole horizon, as given by (\ref{eq:rh_evol}) and (\ref{eq:EE1}), with solution
\be
r_h^3(T_{\SM})=\bar r_h^3\left(1-\frac{T^3_{\SM}}{T_{\SM, R}^3}\right),\qquad \bar r_h^3= \frac{\sqrt{10}C_\rho}{3\pi \sqrt{g_*}}\left(\frac{L T_{\SM,0}}{\eta M_4} \right)^3 T_{\SM,R}^3\,,
\ee
from where we can see that $\bar r_h=r_h(T_{\rm SM}\ll T_{\rm SM,R})$. Then (\ref{eq:valores}) can be written as
\be
\rho_{\DM}(T_{\SM})=3 \eta^2 M_4^2 \left(\frac{r_h(T_{\SM})}{r_b(T_{\SM})} \right)^3\,.
\ee
}
Identifying with \eqref{eq:rhoDMT}, the dark matter abundance predicted by  freeze-in is found to be 
\be
\Omega_{{\rm DM},0} = \frac{\sqrt{10}C_\rho}{3 \pi  \sqrt{g_\star} \eta H_0^2} \left(\frac{T_{\SM, R}T_{\SM,0}}{M_4}\right)^3 \,. \label{eq:Omega_DM}
\ee

\begin{figure}[t]
\begin{center}
 \includegraphics[width=0.4\textwidth]{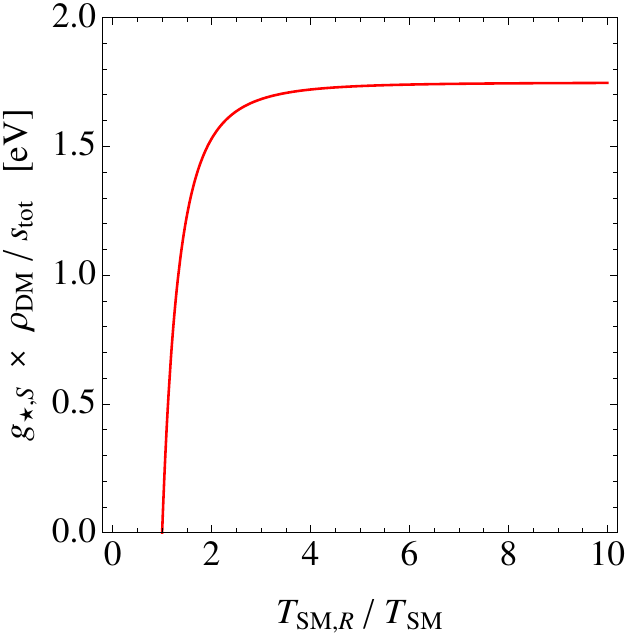}  
\end{center}
     \caption{Energy-to-entropy ratio $\rho_{\rm DM} / s_{\textrm{tot}}$, normalized by $g_{\star,S}$, as a function of $T_{\SM,R} / T_{\SM}$. This quantity freezes  at temperature $T_{\SM}\ll T_{\SM,R}$. 
     }
     \label{fig:TR}
 \end{figure}

Finally, the energy-to-entropy density ratio is given by 
\be
\frac{\rho_{\rm DM}}{s_{\rm tot}}= \frac{45\sqrt{10}C_\rho g_{\star,S}}{2\pi^3 \sqrt{g_\star}}  \frac{T^3_{\SM, R}-T^3_\SM}{ \eta M_4} \label{eq:ratio_result} \,,
\ee
where we used $s_{\rm tot}\approx s_{\SM} = g_{\star,S}\frac{2\pi^2}{45} T^3_\SM$ and  $g_{*,S}$ is the entropy effective number of degrees of freedom in the SM. 
This result is shown in Fig.~\ref{fig:TR}. 

\subsubsection{Discussion}

From \eqref{eq:Omega_DM}, we relate the reheating temperature to the model parameters. Keeping $M_5$ as the free parameter, we find 
\be
T_{\SM,R} = \left[ \frac{3 \pi \Omega_{\rm DM,0} \sqrt{g_*}}{C_\rho \sqrt{10}} \eta H_0^2 \right]^{1/3} \, \frac{M_4}{T_{\SM,0}} \simeq 9.4\times 10^{-10} M_5 \,.
\label{eq:TR}
\ee

Moreover,  considering the bound $M_5\lesssim M_4$ (which corresponds to $\eta\lesssim M_4$) yields for the reheating temperature the upper bound 
\be T_{\SM,R} \lesssim 2.3\times 10^9~{\rm GeV}\,. 
\ee 
This in good agreement with most realistic inflationary models \cite{Cook:2015vqa}.

Finally, the value of the reheating temperature $T_{\textrm{SM},R}$ in (\ref{eq:TR}) needs to satisfy the consistency requirement $\rho_{\rm SM}(T_{\textrm{SM},R})\lesssim \eta^2 M_4^2$, 
 ensuring the validity of the low-energy approximation 
used  in the effective Einstein equation (\ref{eq:D_1_Einstein}). This translates into the mild (sufficient) condition $M_5\gtrsim 10$ GeV, valid for $T_{\SM}<T_{\SM, R}$. 


As expected for freeze-in production, the DM sector is never in thermal equilibrium with the SM as condition (\ref{eq:TR}) guarantees that $\rho_{\rm DM}\ll \rho_{\rm SM}$ at temperatures $T_{\rm SM}\leq T_{\textrm{SM},R}$ (radiation domination). This also justifies one of the assumptions done in
the solving of the evolution equations. The remaining assumption, that $|\Delta\dot\rho_{\rm SM}|\ll H\rho_{\rm SM}$ for any $T_{\rm SM}\leq T_{{\rm SM},R}$, is verified explicitly by using the relation \eqref{eq:TR} in the explicit expression for $\Delta\dot\rho_{\rm DM}$ given in \eqref{eq:c}.

\subsubsection{Emitted gravitons versus dark matter density}
\label{se:comparison}

It is possible to compute the total number of emitted gravitons using the  Boltzmann equation on the brane.  The result is 
$n_{\rm emitted} \propto \frac{T_{\SM,R}^2 T_{\SM}^3}{\eta M_4} $.
On the other hand, the energy density of holographic dark matter obtained in \eqref{eq:rho_DM_result} behaves as  
$\rho_{\rm DM} \propto  \frac{T_{\SM,R}^3 T_{\SM}^3}{\eta M_4} $. 
For non-relativistic matter, one would expect that the ratio of these expressions corresponds to a physical mass scale, i.e. $\frac{\rho}{n}\approx m $. Here, instead, we find that the ratio is proportional to $T_{\SM,R}$, and thus depends on the initial condition.   
This  discrepancy illustrates the fact that $n_{\rm emitted}$ should not be identified as some dark matter number density. 
  Notice that, from the 4D dual perspective,  a strongly-coupled theory  generally does not have particle states.  Thus the notion of number density does not even exist --- the physical quantities relevant for gravity are  energy and pressure.

\section{Phenomenological Constraints}
\label{sec:phenomenology}

In this section we collect cosmological, astrophysical and laboratory
bounds which constrain the holographic dark matter model. 
The model has a single free parameter, see discussion in section in \ref{se:parameters}. Here we express the bounds in terms of the 5D Planck scale $M_5$.

\subsection{Dark Matter Sound Speed}

In section~\ref{sec:DM_freezein} we have determined the speed of sound of holographic dark matter, \eqref{eq:cDM}. This sound speed depends on $r_h^3/r_{b,0}^3$, which is related to $M_5$ by imposing the observed  DM energy density, 
\be
\frac{r_h}{r_{b,0}}\simeq 2.7\times 10^{-4}\left(\frac{\rm GeV}{M_5}\right)^2
\label{eq:rh_bound_bis}
\ee
as shown in~\eqref{eq:rh_bound}. 
The dark matter sound speed is constrained using galactic rotation curves~\cite{Avelino:2015dwa} as
\be
c_\DM \lesssim 10^{-4}\,. 
\ee
Combining \eqref{eq:cDM} and \eqref{eq:rh_bound_bis}, this bound translates as a mild bound on $M_5$, 
\be
M_5\gtrsim 300 \textrm{ MeV}\,.
\ee

\subsection{Supernovae Cooling}

The bulk gravitons can be produced 
 in supernovae (SN)~\cite{Hannestad:2007ys,Lewis:2007ss}.
  If the energy loss due to graviton emission into the bulk were excessive, it would change stellar evolution \cite{Raffelt:1999tx}.
Numerical studies of SN energy losses reveal that the neutrino burst would have been excessively shortened unless the  energy loss rate of the SN core obeys
\be
Q_{\rm SN}\lesssim 3\times 10^{33}\, \textrm{erg cm}^{-3}\textrm{s}^{-1}\simeq 10^{-29} \textrm{ GeV}^5 \,.
\label{eq:boundQ}
\ee
This corresponds to the strongest constraint based on the analysis of SN1987A.

Assuming a SN temperature of $T_{\rm SN} = 30\, \textrm{MeV}$,
the SN energy loss rate is directly given by the graviton emission rate $\Delta\dot{\rho}_{\SM}$  computed  in (\ref{eq:c}), with
\be
Q_{\rm SN}=\Delta\dot{\rho}_{\rm DM} = C_\rho\,\frac{T_{\rm SN}^8}{M_5^3} \,.
\ee
Combining with the bound  (\ref{eq:boundQ}), we obtain the  constraint
\be
M_5\gtrsim 648 \textrm{ TeV}  \,. \label{eq:M5_bound_SN}
\ee
Here we have assumed that the graviton mass gap is negligible with respect to $\eta \ll T_{\rm SN}$, which is verified a posteriori from the value \eqref{eq:M5_bound_SN}.

This bound is stronger than the bounds in \cite{Lewis:2007ss} on unparticles coupled to photons or electrons, as here the graviton continuum is coupled to all SM particles through their energy-momentum tensor.

\subsection{Big Bang Nucleosynthesis (BBN)}

In section \ref{sec:DM_freezein} we have shown that the freeze-in mechanism, upon imposing the observed  DM energy density, implies a relation between the reheating temperature $T_{\SM,R}$ and $M_5$ given by \eqref{eq:TR}. To avoid disturbing the mechanism of BBN, the reheating temperature 
must satisfy the lower bound 
$
T_{\SM,R} \gtrsim  4 \textrm{ MeV} \,.
$
This translates into a bound on $M_5$ as
\be
M_5\gtrsim 4.3\times 10^3 \textrm{ TeV} \,.
\ee

\subsection{Deviation to Newtonian Potential}

The Newtonian potential  at present times in a braneworld model can be deduced from the graviton brane-to-brane propagator~\cite{Fichet:2022xol}. In the linear dilaton braneworld of our focus,  the Newton potential for two particles of masses $m_1$ and $m_2$ at a distance $R$ is given by
\be
V_N(R)=-G_N \frac{m_1 m_2}{R}\left[1+\Delta(R)  \right]  \,,
\ee
where
\be
\Delta(R) \simeq \left\{ \begin{array}{cc}
  \frac{8}{9\pi\eta R}  &\qquad \textrm{if} \quad \eta R \lesssim 1    \\
  \frac{8}{9 \sqrt{3\pi}}  \frac{1}{(\eta R)^{3/2}} e^{- \frac{3}{2} \eta R}  &\qquad \textrm{if} \quad  \eta R \gtrsim 1
  \end{array} \right. \,.
\label{eq:DeltaR}
\ee

Torsion pendulum (fifth-force) experiments~\cite{Hoyle:2004cw,Kapner:2006si,Lee:2020zjt} constrain 
deviations to gravity down to the micrometer scale. 
In the context of our linear dilaton braneworld, we use  the bound $\Delta(R)\lesssim 10^{-2}$ at $R\sim 50\, \mu\textrm{m}$.\,\footnote{Beyond gravity, the  $\propto\frac{1}{r^2}$ fifth force can also be induced by coupling to a CFT operator with conformal dimension $3/2$ \cite{Costantino:2019ixl,Chaffey:2021tmj}. } Applying this bound to \eqref{eq:DeltaR} at  $R \sim 1/\eta$ provides the constraint $\eta \gtrsim 6$~meV.  Translating to $M_5$ we obtain
 \be
 M_5\gtrsim 3 \times 10^5 \textrm{ TeV}\,.
 \label{eq:M5_bound_5f}
 \ee

\subsection{Discussion}

The strongest phenomenological  bound on the 5D Planck mass,  $M_5 \gtrsim 3 \times 10^5$ TeV, comes from the correction to the Newtonian potential. 

The lower bounds on $M_5$ translate as lower bounds on the reheating temperature through~\eqref{eq:TR}. From the fifth force bound of \eqref{eq:M5_bound_5f}, we obtain the mild bound    {$T_{\SM,R} \gtrsim 300$ MeV}. The lowest value allowed is a fairly low reheating temperature, but still consistent with the Standard Cosmological Model~\cite{deSalas:2015glj}. 

Since the graviton couples to the SM with  strength $M_5^{-3/2}$, the above bounds from the fifth force, supernovae, and BBN are all typically stronger than those achievable from present (LHC) or future (FCC-hh) colliders~\cite{Helsens:2019bfw,Harris:2022kls}. In fact, the partonic cross-section production at colliders goes like $\hat\sigma\sim \sqrt{\hat s}/M_5^3$, where $\sqrt{\hat s}$ is the partonic center-of-mass energy. Typically, we can consider for FCC-hh $\sqrt{\hat s}<100$ TeV, which provides, using the bound (\ref{eq:M5_bound_5f}), a production cross-section $\hat\sigma\lesssim 10^{-9}$ fb. For integrated luminosities as high as $10^5\textrm{ fb}^{-1}$ the signal would be below the observability limit.

At the level of laboratory experiments,   future probes of the Newtonian potential, such as optically-levitated sensors~\cite{Moore:2020awi} probing sub-$\mu$m distances,  are likely more promising to further constrain the $M_5$ parameter.

\section{Summary and Outlook}
\label{sec:conclusions}

What is dark matter made of? 
In this work we have developed the idea that dark matter is a fluid/plasma formed in a hidden, strongly-coupled sector. Building on our previous works \cite{Fichet:2022ixi, Fichet:2022xol}, we provide a holographic description of the  DM fluid in terms of  the 5D linear dilaton braneworld. From the 5D viewpoint, the fluid on the brane  originates  from a planar black hole in the bulk. 

We have studied in detail the thermodynamic properties of the holographic fluid, going beyond the  far-horizon approximation  $r_h\ll r_b$. In this limit, the fluid exhibits Hagedorn thermodynamics, hence the free energy vanishes. Going beyond this approximation  is essential  to determine the properties of the black hole phase. We find that the black hole phase is thermodynamically preferred for all radii, and that the associated phase transition is {continuous}. As a consequence,  the cosmological braneworld is always in the black hole phase, as required for the holographic  description of DM.  We also find that the equation of state receives small corrections, and that the dark matter sound speed is very small but nonzero. 

We then present a natural freeze-in mechanism for the emergence of DM in the early universe,  relying solely on the assumption that the inflaton is brane-localized. Under this assumption, the bulk is effectively empty at the time of reheating. After reheating, gravitons are emitted from the brane into the bulk. The energy transferred from the SM thermal bath into the bulk through this process triggers the smooth formation of a horizon. This freeze-in formation of the bulk black hole lasts over one order of magnitude in temperature, after which the {horizon remains unchanged}.

Technically, we present the freeze-in calculation from both the boundary and bulk perspectives, and show that the two descriptions are equivalent. The boundary calculation relies on the effective four-dimensional conservation equation. The bulk calculation mo\-dels the emitted gravitons using a Vaidya-type metric, with the evolution of the horizon following directly from solving the 5D Einstein equations.

After fixing the dark matter abundance to its observed value, the model features a single free parameter. We choose this parameter to be the 5D Planck mass $M_5$. 
We find that the model is constrained by BBN through the reheating temperature and by supernovae due to bulk graviton emission, with $M_5>{\cal O}(10^3)$~TeV. These bounds are dwarfed by torsion pendulum experiments that give {$M_5>{\cal O}(10^5)$~TeV} as leading constraint. A constraint on the DM sound speed also places a modest ${\cal O}({\rm GeV} )$ bound on $M_5$. 

Given these constraints, probing the graviton continuum at present or future colliders does not appear feasible.
More promising prospects may instead come from   future probes of the Newtonian potential, such as optically-levitated sensors~\cite{Moore:2020awi}, which could test the graviton continuum at sub-$\mu$m distances. 

Finally, our model has been used so far to describe the homogeneous dark matter distribution.  
More evolved spherical halo solutions are required to describe the heterogeneous universe.  
Our search for $O(3)$-symmetric solutions is in progress, see  \cite{BlackString} for an encouraging related result.

\begin{acknowledgments}

EM would like to thank the ICTP South American Institute for Fundamental Research (SAIFR), S\~ao Paulo, Brazil, for hospitality and partial financial support during the final stages of this work. The work of EM is supported by the ``Proyectos de Investigaci\'on
  Precompetitivos'' Program of the Plan Propio de Investigaci\'on of
  the University of Granada under grant PP2025PP-18. The work of MQ is
  supported by the grant PID2023-146686NB-C31 funded by
  MI\-CIU/AEI/10.13039/501100011033/ and by FEDER, EU. IFAE is
  partially funded by the CERCA program of the Generalitat de
  Catalunya. SF is supported by grant 2021/10128-0 of FAPESP. 
  
\end{acknowledgments}

\appendix

\section{Computing  Graviton Radiation from the Brane}
\label{app:gravitons}

We present the calculation of the graviton emission rate  from the brane used in section \ref{se:graviton_emission}.

\subsection{Graviton Propagator}

The graviton field is the tensor fluctuation of the 5D metric.  For our purposes it is sufficient to focus on the  transverse traceless component $\tilde  h_{\mu\nu}$, normalized as
\begin{equation}
ds^2 = e^{-2A(z)} \left[ \left(\eta_{\mu\nu} + \frac{2}{M_5^{3/2}} \tilde h_{\mu\nu}(x,z) \right) dx^\mu dx^\nu + dz^2  \right] \,,
\end{equation}
where we are using conformal coordinates.\,\footnote{
The relation between conformal and brane cosmological coordinates is $z = -\frac{1}{\tilde\eta} \log(r / L)$, and the warp factor in conformal coordinates writes $A(z) = \tilde \eta z$. Further details on gauge fixing can be found in \cite{Fichet:2023dju}. }
In these coordinates, the brane at $z = z_b$ partitions the space $\mathcal M$ into two subspaces
$
 \mathcal M^- \equiv \mathcal M \big|_{[z_b,\infty)} $, $\;  \mathcal M^+ \equiv \mathcal M \big|_{(-\infty,z_b]} \,. 
$

We find it convenient to work with the rescaled field $ h_{\mu\nu}(x,z)$ defined by
\begin{equation}
 h_{\mu\nu}(x,z) = e^{-3A(z)/2} \tilde h_{\mu\nu}(x,z) \,,
\end{equation}
whose normalization in the extra dimension is computed as $ \| \tilde h \|^2 \equiv \int dz \, |\tilde h_{\mu\nu}(x,z)|^2$. The corresponding 5D graviton propagator in spacetime $\mathcal M^\pm$ is given by
\be
G^\pm_{\mu\nu;\alpha\beta}(z,z^\prime;p^2) = G^\pm_h(z,z^\prime;p^2) P_{\mu\nu;\alpha\beta} \,,  \quad\quad P_{\mu\nu;\alpha\beta}=g_{\mu\alpha}g_{\nu\beta}-\frac{1}{3}g_{\mu\nu}g_{\alpha\beta}+\mathcal O(p_\mu,\dots)
\ee
where the $\mathcal O(p_\mu)$ terms vanish when $h_{\mu\nu}$ couples to an energy-momentum tensor that is conserved. This will be the case for the process considered here,  for which the coupling is $M_5^{-3/2}h_{\mu\nu}T^{\mu\nu}_{\rm SM}$. The scalar part of the 5D graviton propagator is given by~\footnote{The computation follows the same steps as in \cite{Fichet:2019owx, Megias:2021mgj}. See also \cite{Fichet:2021xfn, Fichet:2023dju} for a  computation of the propagator in the background with one brane and $\nu \ne 1$.}
\begin{eqnarray}
  G_{h}^{+}(z,z^\prime;s) &=& -\frac{1}{C_h} \tilde f_+(z_<)  \left(\tilde f_-(z_>) - \frac{f_-^\prime(z_b)}{f_+^\prime(z_b)} \tilde f_+(z_>) \right)  \,,   \\
G_{h}^{-}(z,z^\prime;s) &=& -\frac{1}{C_h} \left(\tilde f_+(z_<) - \frac{f_+^\prime(z_b)}{f_-^\prime(z_b)} \tilde f_-(z_<) \right) \tilde f_-(z_>)  \,, 
\end{eqnarray}
where $\tilde f_\pm = e^{\pm m_g z\cdot \Delta }$ and $f_\pm = e^{m_g z} \tilde f_\pm$, while
\begin{equation}
\Delta = \sqrt{1 - \frac{s}{m_g^2}} \,, \qquad (s \equiv -p^2) \,,
\end{equation}
and the mass gap is given by
\begin{equation}
m_g = \frac{3}{2} \tilde\eta \,, \qquad \textrm{with} \qquad \tilde\eta = \eta \frac{r_b}{L} \,, \qquad \eta = k \, e^{\bar v_b} \,.
\end{equation}
We are using the notation $z_< = \textrm{min}(z,z^\prime)$ and $z_> = \textrm{max}(z,z^\prime)$. The $C_h$ constant is fixed by the Wronskian of $f_\pm$ in the form
\begin{equation}
C_h \equiv e^{-3A(z)} \left[  f_+^\prime(z) f_-(z) - f_-^\prime(z) f_+(z) \right] =  2 m_g \Delta \,,
\end{equation}
which turns out to be independent of $z$. Finally, the brane-to-brane propagator turns out to be
\be
G_{h}^\pm(z_b,z_b;s)= - \frac{ 1 }{m_g} \frac{1}{\Delta \pm 1} \,.
\ee
Both $\mathcal M^-$ and $\mathcal M^+$ spacetimes feature the same continuum of  modes, starting at $m > m_g$.  $\mathcal M^-$ also have a massless mode corresponding to the physical graviton in 4D (this massless mode is absent in the $\mathcal M^+$). This can be seen by taking the limit of the propagator for $s\to 0$, which gives
\be
G_{h}^-(z_b,z_b;s) \stackrel[ s \to 0 ]{}{\simeq} \frac{2 m_g }{s} \,.
\label{eq:graviton}
\ee
We  get  $G_{h}^- \sim 1/s$, while for $G_{h}^+$ we get a constant behavior. This reproduces the results from \cite{Fichet:2023xbu}. 

In  the following we consider only ${\cal M}^-$ spacetime, hence we focus on the $G^-_h$ propagator.  The spectral function of this propagator is
\begin{equation}
\tilde \rho_h^-(s) \equiv -\frac{1}{\pi} \textrm{Im} \, G_h^{-}(z_b,z_b;s) = 2 m_g \, \delta(s)  + \frac{1}{\pi} \frac{\sqrt{s - m_g^2}}{s} \times \Theta(s - m_g^2) \,. \label{eq:rhoh_m}
\end{equation}
The massless graviton corresponds to the Dirac delta term.

\subsection{Bulk Graviton Emission}

The  scattering amplitude of the graviton exchange process between SM fields $ \varphi + \bar\varphi \to h^*_{\mu\nu} \to \varphi + \bar\varphi  $ is given~by
\be
\mathcal M_{\varphi\varphi\to \varphi\varphi}=\langle 0| h_{\mu\nu}(z_b,x) T^{\mu\nu}_{\varphi}(x) h_{\alpha\beta}(z_b,y) T^{\alpha\beta}_\varphi(y) |0\rangle \,.
\ee
 We write $\varphi=(s,f,V)$, with $s$  for real scalars, $f$ for Dirac fermions and $V$ for real vectors.  

We go to momentum space and focus on the exchange of the graviton continuum. The continuum propagator and corresponding spectral function are given by substracting the zero mode as 
\begin{align}
G_h(z_b,z_b;s )&\equiv G_h^{-}(z_b,z_b;s)-G_h^0(z_b,z_b;s)=\frac{m_g}{s}(\Delta-1) \,, \label{eq:Gh}\\
\rho_h(s)&\equiv \rho_h^-(s)-\rho_h^0(s)=\frac{1}{\pi}\frac{\sqrt{s-m_g^2}}{s}\Theta(s-m_g^2) \,, \label{eq:rhoh}
\end{align}
where $G_h^0$, $\rho_h^0$, is the zero mode contribution in  \eqref{eq:graviton},  \eqref{eq:rhoh_m}.

  The forward amplitude is  written as
\begin{align}
&\mathcal M_{\varphi\varphi\to \varphi\varphi}^{\rm fw} \equiv \mathcal M(\varphi(p_1)\varphi(p_2)\to\varphi(p_1)\varphi(p_2)) = -\frac{1}{M_5^3}  G_{h}(z_b,z_b;s) \times \mathcal T_\varphi \,, \nonumber\\
&\mathcal T_\varphi\equiv \textrm{tr}\,T_\varphi^2-\frac{1}{3}(\textrm{tr}\,T_\varphi)^2 \,,
\label{eq:amplitude}
\end{align}
with $s=-(p_1+p_2)^2$. We have used the identity $ T^{\mu\nu}_\varphi P_{\mu\nu;\alpha\beta}T^{\alpha\beta}_\varphi=\textrm{tr}\,T_\varphi^2-\frac{1}{3}(\textrm{tr}\,T_\varphi)^2  $,  neglected the mass of the SM particles, and averaged over  initial state polarizations.

Finally we  compute the total cross section for $\sigma(\varphi\varphi\to h_{\mu\nu})$ using  the optical theorem, such that
\be
\sigma_{\varphi}\equiv\sigma(\varphi\varphi\to h_{\mu\nu})=\frac{1}{s}\Im \mathcal M^{\rm fw}_{\varphi\varphi\to\varphi\varphi} \,.
\ee
Using Eq.~(\ref{eq:amplitude}) 
we obtain
\begin{equation}
  \sigma_\varphi = \frac{1}{s}   \frac{\sqrt{s - m_g^2} }{s} \, \Theta(s - m_g^2)  \frac{ \mathcal T_\varphi}{M_5^3} =  \frac{c_\varphi}{M_5^3} \sqrt{s - m_g^2} \, \Theta(s - m_g^2)   \,,  \label{eq:sigma}
\end{equation}
with
\be
c_\varphi=(c_s,c_f,c_V)=\left(\frac{1}{12},\frac{1}{16},\frac{1}{4} \right)  \,. \label{eq:c_varphi}
\ee
In the limit $s \gg m_g^2\propto \eta^2$, the local curvature becomes negligible and the result agrees with the one from  flat space~\cite{Giudice:1998ck}.

The energy loss rate is then given by the interaction term of the Boltzmann equation, which can be directly written as
\be
\Delta \dot \rho_{\rm SM} = - \frac{1}{2} \sum_\varphi \kappa_\varphi^2   \int \frac{d^3p_1}{(2\pi)^3} \frac{d^3p_2}{(2\pi)^3}  f_\varphi(E_1) f_\varphi(E_2) (E_1+E_2) \, \sigma_\varphi v_{\textrm{rel}}  \,,
\ee
where $f_\varphi(E)=\frac{g_\varphi}{e^{E/T}\pm 1}$ is the thermal distribution for each of the SM fields  with $g_\varphi$ the multiplicity  and $\kappa_\varphi$  the number of spin degrees of freedom ($\kappa_s=1$ and $\kappa_V = \kappa_f = 2$). This leads to~\eqref{eq:deltarho}.

\bibliographystyle{JHEP}
\bibliography{biblio}

\end{document}